\documentclass[sigconf,nonacm]{acmart}
\usepackage{booktabs}

\AtBeginDocument{%
  \providecommand\BibTeX{{%
    \normalfont B\kern-0.5em{\scshape i\kern-0.25em b}\kern-0.8em\TeX}}}

\setcopyright{acmcopyright}
\copyrightyear{2018}
\acmYear{2018}
\acmDOI{XXXXXXX.XXXXXXX}

\acmPrice{15.00}
\acmISBN{978-1-4503-XXXX-X/18/06}

\begin{document}

\title{No Parameter Left Behind:\\
How Distillation and Model Size Affect Zero-Shot Retrieval
%or\\
%Size Matters for Zero-Shot Retrieval\\
%or\\
%Large Rerankers are Better Generalizable Retrievers
}


\author{Guilherme Moraes Rosa,$^{1,2,3}$ Luiz Bonifacio,$^{1,2}$ Vitor Jeronymo,$^{1,2}$ Hugo Abonizio,$^{1,2}$ Marzieh Fadaee,$^{3}$ Roberto Lotufo,$^{1,2}$ and Rodrigo Nogueira$^{1,2,3}$}
\affiliation{%
  \institution{$^1$NeuralMind, Brazil}
  \institution{$^2$UNICAMP, Brazil}
  \institution{$^3$Zeta Alpha, Netherlands}
 % \streetaddress{8600 Datapoint Drive}
  %\city{Campinas}
  %\state{Sao Paulo}
  \country{}
 % \postcode{78229}
}

\renewcommand{\shortauthors}{Rosa, et al.}

\begin{abstract}

  Recent work has shown that small distilled language models are strong competitors to models that are orders of magnitude larger and slower in a wide range of information retrieval tasks. This has made distilled and dense models, due to latency constraints, the go-to choice for deployment in real-world retrieval applications. In this work, we question this practice by showing that the number of parameters and early query-document interaction play a significant role in the generalization ability of retrieval models. Our experiments show that increasing model size results in marginal gains on in-domain test sets, but much larger gains in new domains never seen during fine-tuning. Furthermore, we show that rerankers largely outperform dense ones of similar size in several tasks. Our largest reranker reaches the state of the art in 12 of the 18 datasets of the Benchmark-IR (BEIR) and surpasses the previous state of the art by 3 average points. Finally, we confirm that in-domain effectiveness is not a good indicator of zero-shot effectiveness. Code is available at \url{https://github.com/guilhermemr04/scaling-zero-shot-retrieval.git}

\end{abstract}

%\begin{CCSXML}
%<ccs2012>
%  <concept_id>10010520.10010553.10010562</concept_id>
%  <concept_desc>Computer systems organization~Embedded systems</concept_desc>
%  <concept_significance>500</concept_significance>
%  <concept_id>10010520.10010575.10010755</concept_id>
%  <concept_desc>Computer systems organization~Redundancy</concept_desc>
%  <concept_significance>300</concept_significance>
%  <concept_id>10010520.10010553.10010554</concept_id>
%  <concept_desc>Computer systems organization~Robotics</concept_desc>
%  <concept_significance>100</concept_significance>
%  <concept_id>10003033.10003083.10003095</concept_id>
%  <concept_desc>Networks~Network reliability</concept_desc>
%  <concept_significance>100</concept_significance>
%</ccs2012>
%\end{CCSXML}

%\ccsdesc[500]{Computer systems organization~Embedded systems}

\keywords{Distillation, Ranking, Dense retrieval, Information Retrieval, Zero-shot Learning}


\maketitle

  \begin{figure}[ht]
  \centering
  \includegraphics[width=8cm]{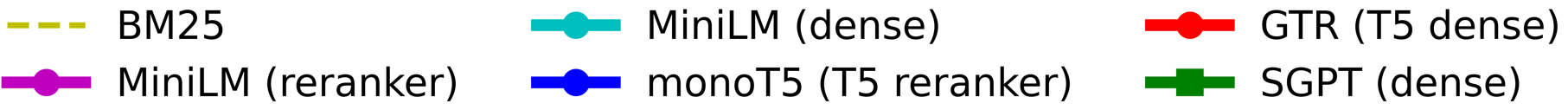}
  \includegraphics[width=8.5cm]{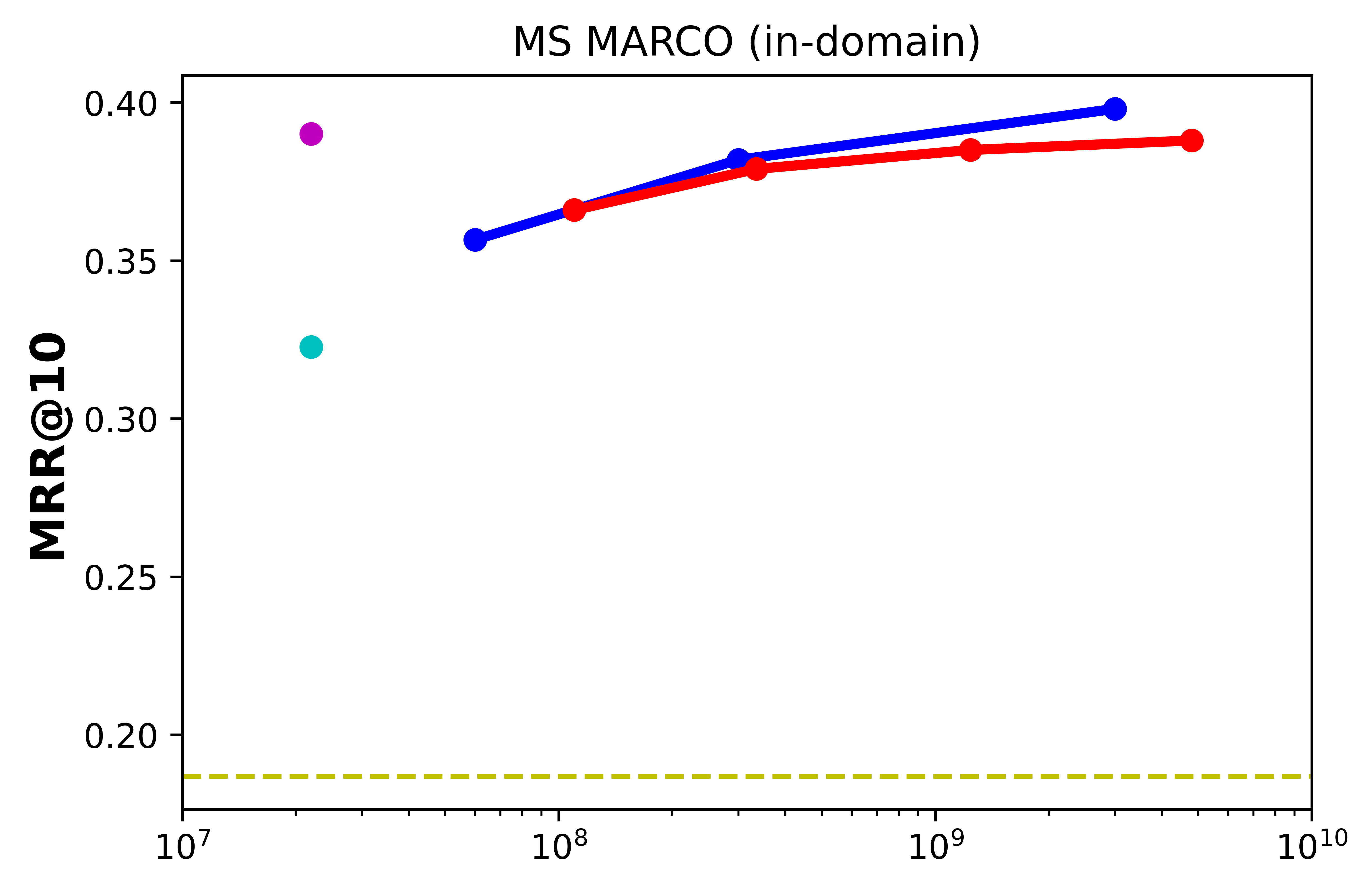}
  \includegraphics[width=8.5cm]{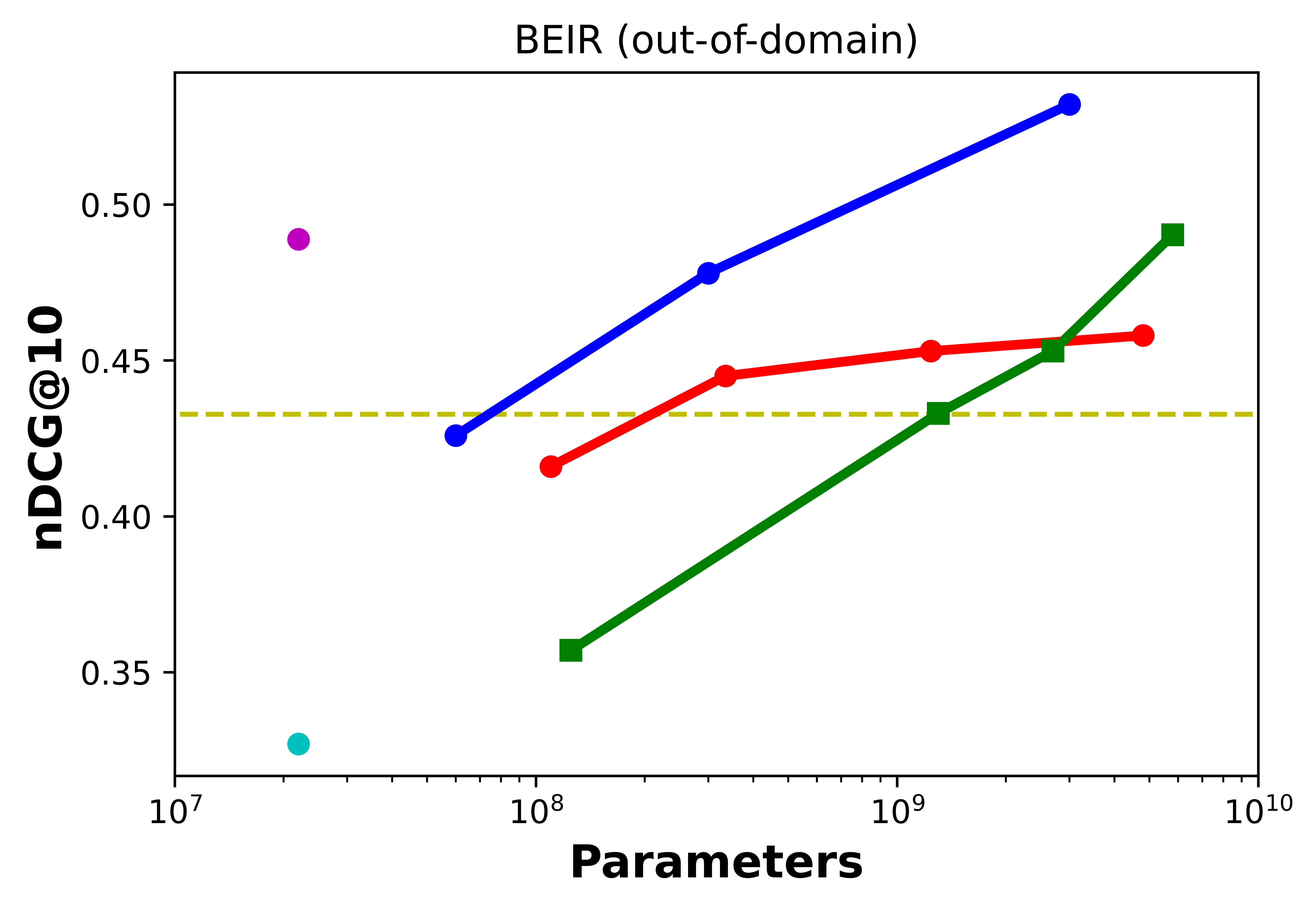}
  \caption{Distilled rerankers such as MiniLM have similar in-domain effectiveness to models 100 times larger (top). However, large rerankers such as monoT5-3B outperform distilled ones and dense models of equivalent size on zero-shot tasks (bottom).}
  \label{fig:graphs_avg}
\end{figure}

\section{Introduction}

Language models have shown remarkable effectiveness in text ranking tasks~\cite{nogueira2019passage,macavaney2019cedr,qu2021rocketqa,gao2021coil,gao-callan-2022-unsupervised}, whose objective is to retrieve an ordered list of texts from a corpus according to their relevance to a given query \cite{lin2020pretrained}. This success suggests great potential for application in commercial search engines \cite{linkedIn}. However, system latency is the main issue to overcome when deploying these language models, as they are computationally intensive and often have to deal with large amounts of data (e.g., collections of documents or web pages). For example, reranking 100 documents for just one query is equivalent to performing 100 predictions in another natural language task (e.g., text classification).

One approach to alleviating this cost is to use knowledge distillation~\cite{hinton2015}. The concept of distillation refers to a set of compression techniques in which a smaller model, called the student, is trained to reproduce the probability distributions predicted by a larger model, called the teacher. This approach is commonly used to decrease model size and control the effectiveness/efficiency trade-off in neural networks, thus reducing inference costs such as latency and memory requirements.
%Previous work has shown that distilled models run faster and occupy less space, while inference can be accelerated by more than an order of magnitude. 
Although knowledge distillation often degrades effectiveness, a potential increase in efficiency can make the trade-off beneficial for certain applications, such as in a text ranking task, since inference needs to be applied over many candidate texts for each query.

It is well known that distilled models fine-tuned on enough annotated data can achieve competitive effectiveness compared to larger models when inferring from similar examples \cite{distilbert,tinybert}. However, in real-world applications, training and inference data often come from different data distributions, as real-world data is generally more diverse than the datasets typically used for fine-tuning \cite{domain2}. This difference can result in poor effectiveness~\cite{domain_adaptation,domain_adaptation2}.%, as language models are still unstable in scenarios involving different data distributions, for example, different textual domains between training and inference data .
%Developing language models capable of overcoming this domain shift is an important challenge. Our objective is to develop models robust enough to generalize well to samples outside of the training data distribution \cite{domain1} and capable of achieving strong zero-shot effectiveness in new domains. An alternative to achieve this goal is the use of larger language models, as it has already been shown that the large number of parameters in these models, up to billions, results in greater generalization capacity and, consequently, superior zero-shot effectiveness compared to smaller models~\cite{NEURIPS2020_1457c0d6,wei2022finetuned,sahn_2021}.

Another approach that offers fast retrieval is dense models, as vector representations of documents can be precomputed prior to retrieval. At retrieval time, only the query vector is computed, and a fast nearest-neighbor search infrastructure is used to retrieve to the most similar document vectors to the query vector. Recent work on dense retrieval mostly compares these models with learned sparse representations~\cite{wang2021gpl,xin2021zero,hofstatter2021efficiently,santhanam2021colbertv2,lu-etal-2021-less,hofstatter2022introducing,thakur2022domain}, while neural rerankers are often ignored due to their computational costs at query time.

However, it is not clear how these retrieval methods compare in zero-shot tasks under different computational budgets.
In this work, we begin by showing that distilled models can achieve competitive effectiveness on an in-domain task with models that are 100x larger. However, we show that in a zero-shot setting, models with a large parameter count have a clear superior effectiveness. Furthermore, our multi-billion parameter reranker achieves state-of-the-art results in the BEIR benchmark, outperforming dense retrieval models of similar size. This suggests that current design choices for dense retrieval methods still have much room for improvement.

\section{Related Work}

Pretrained language models are the state of the art in a wide range of NLP tasks, including text ranking. However, these models often have high computational costs as they consist of hundreds of millions of parameters, which poses challenges for deployment in real-world applications mainly due to latency constraints. To address this challenge \citet{distilbert} proposed an approach to pretrain smaller language models, called DistilBERT, that can be fine-tuned on a wide range of tasks, similar to larger models. They were able to reduce the size of the model by 40\% while retaining 97\% of its effectiveness and being 60\% faster. \citet{tinybert} proposed a new knowledge distillation method specially designed for transformer-based models to speed up inference and reduce model size while maintaining accuracy. The method performs the distillation of knowledge in the pretraining and task-specific learning stages to ensure that the model can capture the general-domain as well as the task-specific knowledge in BERT. Furthermore, they also presented TinyBERT, an effective distilled model that achieved more than 96.8\% of its teacher's effectiveness on GLUE benchmark \cite{glue}, while being 7.5x smaller and 9.4x faster during inference. 

\citet{mobileBERT} proposed MobileBERT, a distilled version of BERT-large that can be applied to many downstream tasks through a simple fine-tuning procedure. The authors showed that MobileBERT is 4.3× smaller and 5.5× faster than BERT-base, while achieving competitive results on various benchmarks such as GLUE \cite{glue} and SQuAD \cite{squad2018}. \citet{minilm} presented an effective and simple approach to compress large pretrained transformer models, named deep self-attention distillation. They proposed to distill the self-attention module of the last transformer layer of the teacher model. The results showed that the student model outperforms state-of-the-art baselines while maintaining more than 99\% of the accuracy on SQuAD and several GLUE benchmark tasks, and using only 50\% of the teacher model parameters. \citet{xtremedistil} proposed a distillation approach that surpassed all the strategies employed in previous works. They demonstrated that this novel approach leads to massive compression of models, up to 35x in terms of parameters and 51x in terms of latency, while retaining 95\% of the score for a named entity recognition task in 41 languages.

Zero-shot models demonstrated state-of-the-art effectiveness on various language tasks. For example, \citet{nogueira2020document} proposed monoT5, a novel adaptation of a pretrained sequence-to-sequence language model designed for the text ranking task. The model is fine-tuned to generate a score that measures a document's relevance to a given query and has achieved state-of-the-art zero-shot results on TREC 2004 Robust Track \cite{trec2004}. Furthermore, this approach significantly outperformed fine-tuned models in data scarcity scenarios. \citet{pradeep5h2oloo} used a monoT5 model trained only on a general domain text ranking dataset to reach the best or second best effectiveness on multiple tasks in specific domains, e.g. the medical domain~\cite{Roberts2019OverviewOT}, including documents about COVID-19~\cite{zhang2020rapidly}. \citet{icail_2021}, using the same T5-based model developed for text ranking, showed for the legal domain that a zero-shot model can perform better than models fine-tuned on the task itself. They argued that in a limited annotated data scenario, a zero-shot model fine-tuned only on a large general domain dataset may generalize better on unseen data than models fine-tuned on a small domain-specific dataset. \citet{sahn_2021} created a system to map any NLP task to prompts written in natural language and converted many supervised datasets into that format. Next, they fine-tuned an 11 billion-parameter T5 model on that combined dataset and found that the model achieves great zero-shot effectiveness across multiple unseen datasets, outperforming models up to 16x its size. \citet{lin_2021} fine-tuned a 7.5 billion-parameters multilingual model in a number of languages to study its zero-shot learning capabilities across multiple tasks. The model achieved state-of-the-art results in over 20 languages, outperforming even some larger language models on various tasks such as natural language inference. 

\begin{table*}[]
\centering%\centering\resizebox{0.5\textwidth}{!}{
\begin{tabular}{lc|cccc|ccc}
\toprule
 & & \multicolumn{4}{c|}{Reranking top 1000 docs from BM25} & \multicolumn{3}{c}{Dense Models}\\
 & \textbf{BM25} & \textbf{MiniLM$^1$} & \multicolumn{3}{c|}{\textbf{monoT5}} & \textbf{ColBERT-v2$^1$} & \textbf{GTR} & \textbf{SGPT}   \\
\textbf{Parameters} & - & 22M & 60M & 220M & 3B & 110M & 4.8B & 5.8B \\
\midrule
MS MARCO & 0.1870 & 0.3901 
& 0.3566 & 0.3810 & 0.3980 & 0.3970 & 0.3880 & -   \\ 
\midrule
TREC-COVID & 0.5947 & 0.7188 & 0.6928 &  0.7775 & 0.7948 & 0.7380 & 0.5010 & \textbf{0.8730}  \\
NFCorpus & 0.3218 & 0.3501 & 0.3180 & 0.3570 & \textbf{0.3837} & 0.3380 & 0.3420 & 0.3630 \\  
BioASQ & 0.5224 & 0.5335 & 0.4880 & 0.5240 & \textbf{0.5740} & - & 0.3240 & 0.4130  \\
%\midrule
Natural Questions & 0.3055 & 0.5525 
& 0.4733 & 0.5674 & \textbf{0.6334} & 0.5620 & 0.5680 & 0.5240 \\
HotpotQA & 0.6330 & 0.7324 
& 0.5996 & 0.6950 & \textbf{0.7589} & 0.6670 & 0.5990 & 0.5930 \\
FEVER & 0.6513 & 0.8180 
& 0.7191 & 0.8018 & \textbf{0.8495} & 0.7850 & 0.7400 & 0.7830 \\
Climate-FEVER & 0.1651 & 0.2555 & 0.2116 & 0.2451 & 0.2802 & 0.1760 & 0.2670 & \textbf{0.3050} \\
DBPedia & 0.3180 & 0.4652 
& 0.3437 & 0.4195 & \textbf{0.4777} & 0.4460 & 0.4080 & 0.3990  \\
%\midrule
TREC-NEWS & 0.3952 & 0.4464 
& 0.3848 & 0.4475 & 0.4727 & - & 0.3460 & \textbf{0.4810} \\
Robust04 & 0.4485 & 0.4801 
& 0.4222 & 0.5016 & \textbf{0.6148} & - & 0.5060 & 0.5140 \\
%\midrule
ArguAna & 0.5302 & 0.5021 
& 0.1274 & 0.1946 & 0.3803 & 0.4630 & \textbf{0.5400} & 0.5140 \\
Touché-2020 & \textbf{0.4422} & 0.2812 & 0.2643 & 0.2773 & 0.2995 & 0.2630 & 0.2560 & 0.2540 \\
%\midrule
CQADupStack & 0.2788 & 0.3611 & 0.3474 & 0.3808 & \textbf{0.4155} & - & 0.3990 & 0.3810 \\
Quora & 0.7886 & 0.8037 & 0.8259 & 0.8230 & 0.8407 & - & \textbf{0.8920} & 0.8460 \\
%\midrule
SCIDOCS & 0.1490 & 0.1629 
& 0.1436 & 0.1649 & \textbf{0.1970} & 0.1540 & 0.1610 & \textbf{0.1970}  \\
SciFact & 0.6789 & 0.6812 
& 0.6963 & 0.7356 & \textbf{0.7773} & 0.6930 & 0.6620 & 0.7470 \\
%\midrule
FiQA-2018 & 0.2361 & 0.3599 
& 0.3377 & 0.4136 & \textbf{0.5137} & 0.3560 & 0.4670 & 0.3720  \\
%\midrule
Signal-1M (RT) & \textbf{0.3304} & 0.2964 & 0.2711 & 0.2771 & 0.3140 & - & 0.2730 & 0.2670 \\
\midrule
Avg (excl. MARCO) & 0.4328 & 0.4889 & 0.4259 & 0.4780 & \textbf{0.5321} & - & 0.4580 & 0.4903  \\
Improvement over BM25 & - & 0.0561 & -0.0069 & 0.0452 & \textbf{0.0993} & - & 0.0252 & 0.0575  \\
\bottomrule
\end{tabular}
%}
\caption{Results (nDCG@10) on the BEIR benchmark. We use MRR@10 for MS MARCO. All results except MS MARCO are zero-shot. $^{1}$Distilled models.}
\label{tab:main_results}
\end{table*}

Additionally, larger language models demonstrated stronger zero-shot and few-shot capabilities compared to smaller models. \citet{NEURIPS2020_1457c0d6} presented GPT-3, a language model with 175 billion parameters. The authors evaluated the model's ability to learn from a few examples and demonstrated that scaling up the number of parameters greatly improves zero-shot and few-shot effectiveness. \citet{wei2022finetuned} presented \textit{instruction tuning}, a novel method to improve the zero-shot learning capacity of scaled language models by providing task-specific instructions using natural language. This method substantially improved zero-shot effectiveness and the model surpassed the GPT-3 on several datasets. \citet{metalmm} proposed \textit{meta-tuning}, a method for optimizing zero-shot learning of scaled language models by fine-tuning on a collection of datasets in a question answering format. The model surpassed similar size language models, including state-of-the-art ones. Furthermore, recent results suggest that we can still improve effectiveness by following this trend of increasing model size. To understand the impact of scaling in few-shot learning, \citet{PALM} introduced the Pathways Language Model (PaLM), a 540 billion parameter language model that has achieved state-of-the-art few-shot results in hundreds of language tasks, also outperforming multiple fine-tuned state-of-the-art models.

Domain adaptation is often necessary when applying NLP models to real-world scenarios as it deals with the challenge of transferring knowledge from a source domain to a different domain at the time of inference, i.e., the model must be able to extrapolate well across different domains. Although there is extensive study on the effectiveness of NLP models on in-domain data, we still know little about their ability to extrapolate to out-of-domain datasets. To better understand this issue, \citet{dense_2022} investigated the extrapolation effectiveness of dense retrieval models. They found that dense retrieval models can achieve competitive effectiveness compared to reranker models on in-domain tasks, but extrapolate substantially worse to new domains. Their results also indicated that pretraining the model on target domain data improves the extrapolation capability. \citet{dr_review} presented an extensive and detailed examination of the zero-shot capability of dense retrieval models to better understand them. The authors analyzed key factors related to zero-shot retrieval effectiveness, such as source dataset, potential bias in target dataset and also reviewed and compared different dense retrieval models. \citet{dual_2021} presented Generalizable T5-based dense Retrievers (GTR), a model designed to deal with the problem of poor effectiveness outside the domain of previous dual encoder models. The authors found that increasing the size of the model brings a significant improvement in out-of-domain generalization. The model outperformed existing sparse and dense models in a variety of retrieval tasks, while being fine-tuned on only 10\% of the MS MARCO dataset. \citet{SGPT} argued that although GPT transformers are the largest language models available, information retrieval tasks are dominated by encoder-only or encoder-decoder transformers. To change this scenario, \citet{SGPT} presented the SGPT model, a 5.8 billion parameter model that achieved state-of-the-art results on several out-of-domain information retrieval datasets and also outperformed much larger models with up to 175 billion parameters. %The author also demonstrated that effectiveness can be improved by adapting the prompt used as input to the model.

\section{Experiments} 

In our experiments, we evaluate three different models: BM25 \cite{lin2021pyserini}, MiniLM-L6~\cite{minilm} and monoT5~\cite{nogueira2020document} in three different sizes: 60M (small), 220M (base) and 3B parameters.
%BM25 is a retrieval algorithm that assigns scores to documents based on the terms of a query that appear in it.
We use BM25 implemented in Pyserini~\cite{lin2021pyserini}, a Python library for information retrieval. MiniLM is a distilled reranker proposed by \citet{minilm} that achieves competitive effectiveness compared to state-of-the-art models in multiple retrieval tasks.
MonoT5 is an adaptation of the T5 model \cite{raffel2020t5} proposed by Nogueira et al. \cite{nogueira2020document}. The model takes a query-document pair as input and is designed to generate a probability score that quantifies the relevance between them. All three of our transformer models are fine-tuned on MS MARCO \cite{marco} and available at Hugging Face.\footnote{\url{https://huggingface.co/cross-encoder/ms-marco-MiniLM-L-6-v2},\\\url{https://huggingface.co/castorini/monot5-base-msmarco-10k},\\ \url{https://huggingface.co/castorini/monot5-3B-msmarco-10k}} MS MARCO is a large-scale dataset for passage ranking task, consisting of 8.8 million passages taken from the Bing search engine. The fine-tuning procedures we use in this work are described in detail in \citet{reimers-2019-sentence-bert} and \citet{nogueira2020document} for MiniLM and monoT5, respectively. At inference time, rerankers sort a list of 1000 texts retrieved by BM25 according to their relevance to a query.

We evaluate the zero-shot effectiveness of our models on the BEIR benchmark~\citet{beir}, which consists of 18 publicly available datasets from different domains, such as web, biomedical and financial. With the exception of the MS MARCO dataset, all of our results are zero-shot as our models are fine-tuned on MS MARCO and directly evaluated on other datasets.

\section{Results}

Results in Table~\ref{tab:main_results} show that while MiniLM achieves excellent results on MS MARCO, including competitive effectiveness to a much larger model such as monoT5-3B, the distilled model underperforms in comparison to monoT5-3B on almost all datasets outside the domain seen during fine-tuning. For example, on the FiQA dataset, which comprises text in the financial domain, MiniLM performs nearly 16 points below our largest reranker. Other significant results include the experiments on the Scifact and Natural Questions datasets, in which the MiniLM is also outperformed by monoT5-3B by almost 10 and 8 points, respectively. Nonetheless, average results show that MiniLM achieves competitive effectiveness compared to monoT5-base, which has 10 times more parameters, suggesting that MiniLM is able to retain some degree of generalization capability.

As shown in Figure~\ref{fig:graphs_avg}, the average results of monoT5-3B and SGPT-5.8B demonstrate that strong zero-shot effectiveness in new text domains can be achieved by increasing the number of model parameters and without fine-tuning on in-domain data. However, our 3 billion parameter reranker model outperformed SGPT, a state-of-the-art dense retrieval model almost twice as large, and ColBERTv2~\cite{santhanam2021colbertv2}, a distilled dense model.
Another interesting comparison is that monoT5 and GTR use the same underlying model, T5, but the reranker has better effectiveness than the dense retriever. GTR's effectiveness increases at a slower rate with more parameters in comparison to monoT5's. We suspect that this ``saturation'' is due to the decision to keep the size of document and query vectors fixed to 768 dimensions as the model size increases. Results for the MiniLM model reveal a similar behavior: the MiniLM reranker outperforms its dense version on both MS MARCO and BEIR.
These results suggest that early query-document interactions in the reranker architecture and a high number of parameters (and hidden dimensions) have a positive impact on zero-shot effectiveness.

\subsection{Latency}

A major drawback of rerankers in comparison to dense retrievers is their latency, as the former requires one inference pass of a potentially large neural network per query-document pair, while dense retrievers can compute the vector representation of documents prior to retrieval, thus saving considerable computation at query time.

In Table~\ref{tab:latency} we compare effectiveness (nDCG@10 on BEIR) with efficiency (seconds/query) of rerankers of different sizes. Our experiments were performed on an A100 GPU with 40GB. Despite receiving only 50 documents per query, monoT5-3B consistently outperforms all other smaller models, while performing almost as efficiently as MiniLM, a distilled model more than 100 times smaller. %For small rerankers, decreasing the number of reranked documents is not an effective strategy to reduce latency. For exemple, reranking 50 documents with miniLM takes XXX seconds, which is only a X\% faster than reranking 1000 documents. This is because, for small models, we have to use large batch sizes to efficiently use the GPU.

\begin{table}[]
\centering%\centering\resizebox{0.5\textwidth}{!}{
\begin{tabular}{lrrrr}
\toprule
\textbf{Reranker} & \textbf{Params} & \textbf{docs/query} & \textbf{nDCG@10} & \textbf{s/query} \\
\midrule
MiniLM & 22M & 1000 & 0.4889 & 0.64 \\
monoT5 & 60M & 1000 & 0.4259 & 2.02 \\
monoT5 & 220M & 1000 & 0.4780 & 3.00 \\
monoT5 & 3B & 1000 & 0.5321 & 14.63 \\
monoT5 & 3B & 50 & 0.5152 & 0.73 \\
\bottomrule
\end{tabular}
%}
\caption{Rerankers effectiveness (nDCG@10) vs efficiency (s/query) on BEIR. Candidate documents are from BM25.}
\label{tab:latency}
\end{table}

\section{Conclusion}

In this work we studied how distillation and parameter count influence the zero-shot effectiveness of neural retrievers. We begin by showing that in-domain effectiveness, i.e., when retrievers are finetuned and evaluated on the same dataset such as MS MARCO, is not a good proxy for zero-shot effectiveness, which corroborates recent claims by Lin et al.~\cite{lin2022fostering} and Gupta et al.~\cite{gupta2022survivorship} and Zhan et al~\cite{dense_2022}.

Furthermore, we show that a distilled reranker has better zero-shot effectiveness than much larger non-distilled rerankers, which is an important and desirable feature of deployed models. However, our largest reranker significantly outperforms smaller rerankers and achieves a new state of the art across almost all datasets used in our zero-shot experiments. This suggests that a large number of parameters may play a significant role in the generalization capability of pretrained language models.

Lastly, our results confirm the findings of Zhan et al.~\cite{dense_2022}, who show that dense retrievers have poorer generalization than rerankers to new domains. Our study, however, shows that this is the case across a wide range of model sizes, suggesting that much work needs to be done in dense retrieval methods.

\bibliographystyle{ACM-Reference-Format}
\bibliography{sample-base}


\begin{thebibliography}{48}

%%% the Web address.

\ifx \showCODEN    \undefined \def \showCODEN     #1{\unskip}     \fi
\ifx \showDOI      \undefined \def \showDOI       #1{#1}\fi
\ifx \showISBNx    \undefined \def \showISBNx     #1{\unskip}     \fi
\ifx \showISBNxiii \undefined \def \showISBNxiii  #1{\unskip}     \fi
\ifx \showISSN     \undefined \def \showISSN      #1{\unskip}     \fi
\ifx \showLCCN     \undefined \def \showLCCN      #1{\unskip}     \fi
\ifx \shownote     \undefined \def \shownote      #1{#1}          \fi
\ifx \showarticletitle \undefined \def \showarticletitle #1{#1}   \fi
\ifx \showURL      \undefined \def \showURL       {\relax}        \fi
% invisible to TeX
\providecommand\bibfield[2]{#2}
\providecommand\bibinfo[2]{#2}
\providecommand\natexlab[1]{#1}
\providecommand\showeprint[2][]{arXiv:#2}

\bibitem[Bajaj et~al\mbox{.}(2018)]%
        {marco}
\bibfield{author}{\bibinfo{person}{Payal Bajaj}, \bibinfo{person}{Daniel
  Campos}, \bibinfo{person}{Nick Craswell}, \bibinfo{person}{Li Deng},
  \bibinfo{person}{Jianfeng Gao}, \bibinfo{person}{Xiaodong Liu},
  \bibinfo{person}{Rangan Majumder}, \bibinfo{person}{Andrew McNamara},
  \bibinfo{person}{Bhaskar Mitra}, \bibinfo{person}{Tri Nguyen},
  \bibinfo{person}{Mir Rosenber}, \bibinfo{person}{Xia Song},
  \bibinfo{person}{Alina Stoica}, \bibinfo{person}{Saurabh Tiwary}, {and}
  \bibinfo{person}{Tong Wang}.} \bibinfo{year}{2018}\natexlab{}.
\newblock \showarticletitle{MS MARCO: A Human Generated MAchine Reading
  Comprehension Dataset}.
\newblock \bibinfo{journal}{\emph{arXiv preprint arXiv:1611.09268}}
  (\bibinfo{year}{2018}).
\newblock


\bibitem[Ben-David et~al\mbox{.}(2021)]%
        {domain_adaptation2}
\bibfield{author}{\bibinfo{person}{Eyal Ben-David}, \bibinfo{person}{Nadav
  Oved}, {and} \bibinfo{person}{Roi Reichart}.}
  \bibinfo{year}{2021}\natexlab{}.
\newblock \showarticletitle{PADA: Example-based Prompt Learning for on-the-fly
  Adaptation to Unseen Domains}.
\newblock \bibinfo{journal}{\emph{arXiv preprint arXiv:2102.12206}}
  (\bibinfo{year}{2021}).
\newblock


\bibitem[Brown et~al\mbox{.}(2020)]%
        {NEURIPS2020_1457c0d6}
\bibfield{author}{\bibinfo{person}{Tom Brown}, \bibinfo{person}{Benjamin Mann},
  \bibinfo{person}{Nick Ryder}, \bibinfo{person}{Melanie Subbiah},
  \bibinfo{person}{Jared~D Kaplan}, \bibinfo{person}{Prafulla Dhariwal},
  \bibinfo{person}{Arvind Neelakantan}, \bibinfo{person}{Pranav Shyam},
  \bibinfo{person}{Girish Sastry}, \bibinfo{person}{Amanda Askell},
  \bibinfo{person}{Sandhini Agarwal}, \bibinfo{person}{Ariel Herbert-Voss},
  \bibinfo{person}{Gretchen Krueger}, \bibinfo{person}{Tom Henighan},
  \bibinfo{person}{Rewon Child}, \bibinfo{person}{Aditya Ramesh},
  \bibinfo{person}{Daniel Ziegler}, \bibinfo{person}{Jeffrey Wu},
  \bibinfo{person}{Clemens Winter}, \bibinfo{person}{Chris Hesse},
  \bibinfo{person}{Mark Chen}, \bibinfo{person}{Eric Sigler},
  \bibinfo{person}{Mateusz Litwin}, \bibinfo{person}{Scott Gray},
  \bibinfo{person}{Benjamin Chess}, \bibinfo{person}{Jack Clark},
  \bibinfo{person}{Christopher Berner}, \bibinfo{person}{Sam McCandlish},
  \bibinfo{person}{Alec Radford}, \bibinfo{person}{Ilya Sutskever}, {and}
  \bibinfo{person}{Dario Amodei}.} \bibinfo{year}{2020}\natexlab{}.
\newblock \showarticletitle{Language Models are Few-Shot Learners}. In
  \bibinfo{booktitle}{\emph{Advances in Neural Information Processing
  Systems}}, \bibfield{editor}{\bibinfo{person}{H.~Larochelle},
  \bibinfo{person}{M.~Ranzato}, \bibinfo{person}{R.~Hadsell},
  \bibinfo{person}{M.~F. Balcan}, {and} \bibinfo{person}{H.~Lin}} (Eds.),
  Vol.~\bibinfo{volume}{33}. \bibinfo{publisher}{Curran Associates, Inc.},
  \bibinfo{pages}{1877--1901}.
\newblock
\urldef\tempurl%
\url{https://proceedings.neurips.cc/paper/2020/file/1457c0d6bfcb4967418bfb8ac142f64a-Paper.pdf}
\showURL{%
\tempurl}


\bibitem[Chowdhery et~al\mbox{.}(2022)]%
        {PALM}
\bibfield{author}{\bibinfo{person}{Aakanksha Chowdhery},
  \bibinfo{person}{Sharan Narang}, \bibinfo{person}{Jacob Devlin},
  \bibinfo{person}{Maarten Bosma}, \bibinfo{person}{Gaurav Mishra},
  \bibinfo{person}{Adam Roberts}, \bibinfo{person}{Paul Barham},
  \bibinfo{person}{Hyung~Won Chung}, \bibinfo{person}{Charles Sutton},
  \bibinfo{person}{Sebastian Gehrmann}, \bibinfo{person}{Parker Schuh},
  \bibinfo{person}{Kensen Shi}, \bibinfo{person}{Sasha Tsvyashchenko},
  \bibinfo{person}{Joshua Maynez}, \bibinfo{person}{Abhishek Rao},
  \bibinfo{person}{Parker Barnes}, \bibinfo{person}{Yi Tay},
  \bibinfo{person}{Noam Shazeer}, \bibinfo{person}{Vinodkumar Prabhakaran},
  \bibinfo{person}{Emily Reif}, \bibinfo{person}{Nan Du}, \bibinfo{person}{Ben
  Hutchinson}, \bibinfo{person}{Reiner Pope}, \bibinfo{person}{James Bradbury},
  \bibinfo{person}{Jacob Austin}, \bibinfo{person}{Michael Isard},
  \bibinfo{person}{Guy Gur-Ari}, \bibinfo{person}{Pengcheng Yin},
  \bibinfo{person}{Toju Duke}, \bibinfo{person}{Anselm Levskaya},
  \bibinfo{person}{Sanjay Ghemawat}, \bibinfo{person}{Sunipa Dev},
  \bibinfo{person}{Henryk Michalewski}, \bibinfo{person}{Xavier Garcia},
  \bibinfo{person}{Vedant Misra}, \bibinfo{person}{Kevin Robinson},
  \bibinfo{person}{Liam Fedus}, \bibinfo{person}{Denny Zhou},
  \bibinfo{person}{Daphne Ippolito}, \bibinfo{person}{David Luan},
  \bibinfo{person}{Hyeontaek Lim}, \bibinfo{person}{Barret Zoph},
  \bibinfo{person}{Alexander Spiridonov}, \bibinfo{person}{Ryan Sepassi},
  \bibinfo{person}{David Dohan}, \bibinfo{person}{Shivani Agrawal},
  \bibinfo{person}{Mark Omernick}, \bibinfo{person}{Andrew~M. Dai},
  \bibinfo{person}{Thanumalayan~Sankaranarayana Pillai}, \bibinfo{person}{Marie
  Pellat}, \bibinfo{person}{Aitor Lewkowycz}, \bibinfo{person}{Erica Moreira},
  \bibinfo{person}{Rewon Child}, \bibinfo{person}{Oleksandr Polozov},
  \bibinfo{person}{Katherine Lee}, \bibinfo{person}{Zongwei Zhou},
  \bibinfo{person}{Xuezhi Wang}, \bibinfo{person}{Brennan Saeta},
  \bibinfo{person}{Mark Diaz}, \bibinfo{person}{Orhan Firat},
  \bibinfo{person}{Michele Catasta}, \bibinfo{person}{Jason Wei},
  \bibinfo{person}{Kathy Meier-Hellstern}, \bibinfo{person}{Douglas Eck},
  \bibinfo{person}{Jeff Dean}, \bibinfo{person}{Slav Petrov}, {and}
  \bibinfo{person}{Noah Fiedel}.} \bibinfo{year}{2022}\natexlab{}.
\newblock \showarticletitle{PaLM: Scaling Language Modeling with Pathways}.
\newblock \bibinfo{journal}{\emph{arXiv preprint arXiv:2204.02311}}
  (\bibinfo{year}{2022}).
\newblock


\bibitem[Farahani et~al\mbox{.}(2020)]%
        {domain2}
\bibfield{author}{\bibinfo{person}{Abolfazl Farahani}, \bibinfo{person}{Sahar
  Voghoei}, \bibinfo{person}{Khaled Rasheed}, {and} \bibinfo{person}{Hamid~R.
  Arabnia}.} \bibinfo{year}{2020}\natexlab{}.
\newblock \showarticletitle{A Brief Review of Domain Adaptation}.
\newblock \bibinfo{journal}{\emph{arXiv preprint arXiv:2010.03978}}
  (\bibinfo{year}{2020}).
\newblock


\bibitem[Gao and Callan(2022)]%
        {gao-callan-2022-unsupervised}
\bibfield{author}{\bibinfo{person}{Luyu Gao} {and} \bibinfo{person}{Jamie
  Callan}.} \bibinfo{year}{2022}\natexlab{}.
\newblock \showarticletitle{Unsupervised Corpus Aware Language Model
  Pre-training for Dense Passage Retrieval}. In
  \bibinfo{booktitle}{\emph{Proceedings of the 60th Annual Meeting of the
  Association for Computational Linguistics (Volume 1: Long Papers)}}.
  \bibinfo{publisher}{Association for Computational Linguistics},
  \bibinfo{address}{Dublin, Ireland}, \bibinfo{pages}{2843--2853}.
\newblock
\urldef\tempurl%
\url{https://aclanthology.org/2022.acl-long.203}
\showURL{%
\tempurl}


\bibitem[Gao et~al\mbox{.}(2021)]%
        {gao2021coil}
\bibfield{author}{\bibinfo{person}{Luyu Gao}, \bibinfo{person}{Zhuyun Dai},
  {and} \bibinfo{person}{Jamie Callan}.} \bibinfo{year}{2021}\natexlab{}.
\newblock \showarticletitle{COIL: Revisit Exact Lexical Match in Information
  Retrieval with Contextualized Inverted List}. In
  \bibinfo{booktitle}{\emph{Proceedings of the 2021 Conference of the North
  American Chapter of the Association for Computational Linguistics: Human
  Language Technologies}}. \bibinfo{pages}{3030--3042}.
\newblock


\bibitem[Guo et~al\mbox{.}(2021)]%
        {linkedIn}
\bibfield{author}{\bibinfo{person}{Weiwei Guo}, \bibinfo{person}{Xiaowei Liu},
  \bibinfo{person}{Sida Wang}, \bibinfo{person}{Michaeel Kazi},
  \bibinfo{person}{Zhoutong Fu}, \bibinfo{person}{Huiji Gao},
  \bibinfo{person}{Jun Jia}, \bibinfo{person}{Liang Zhang}, {and}
  \bibinfo{person}{Bo Long}.} \bibinfo{year}{2021}\natexlab{}.
\newblock \showarticletitle{Deep Natural Language Processing for LinkedIn
  Search Systems}.
\newblock \bibinfo{journal}{\emph{arXiv preprint arXiv:2108.08252}}
  (\bibinfo{year}{2021}).
\newblock


\bibitem[Gupta and MacAvaney(2022)]%
        {gupta2022survivorship}
\bibfield{author}{\bibinfo{person}{Prashansa Gupta} {and} \bibinfo{person}{Sean
  MacAvaney}.} \bibinfo{year}{2022}\natexlab{}.
\newblock \showarticletitle{On Survivorship Bias in MS MARCO}.
\newblock \bibinfo{journal}{\emph{arXiv preprint arXiv:2204.12852}}
  (\bibinfo{year}{2022}).
\newblock


\bibitem[Hinton et~al\mbox{.}(2015)]%
        {hinton2015}
\bibfield{author}{\bibinfo{person}{Geoffrey Hinton}, \bibinfo{person}{Oriol
  Vinyals}, {and} \bibinfo{person}{Jeff Dean}.}
  \bibinfo{year}{2015}\natexlab{}.
\newblock \showarticletitle{Distilling the Knowledge in a Neural Network}.
\newblock \bibinfo{journal}{\emph{arXiv preprint arXiv:1503.02531}}
  (\bibinfo{year}{2015}).
\newblock


\bibitem[Hofst{\"a}tter et~al\mbox{.}(2022)]%
        {hofstatter2022introducing}
\bibfield{author}{\bibinfo{person}{Sebastian Hofst{\"a}tter},
  \bibinfo{person}{Omar Khattab}, \bibinfo{person}{Sophia Althammer},
  \bibinfo{person}{Mete Sertkan}, {and} \bibinfo{person}{Allan Hanbury}.}
  \bibinfo{year}{2022}\natexlab{}.
\newblock \showarticletitle{Introducing Neural Bag of Whole-Words with
  ColBERTer: Contextualized Late Interactions using Enhanced Reduction}.
\newblock \bibinfo{journal}{\emph{arXiv preprint arXiv:2203.13088}}
  (\bibinfo{year}{2022}).
\newblock


\bibitem[Hofst{\"a}tter et~al\mbox{.}(2021)]%
        {hofstatter2021efficiently}
\bibfield{author}{\bibinfo{person}{Sebastian Hofst{\"a}tter},
  \bibinfo{person}{Sheng-Chieh Lin}, \bibinfo{person}{Jheng-Hong Yang},
  \bibinfo{person}{Jimmy Lin}, {and} \bibinfo{person}{Allan Hanbury}.}
  \bibinfo{year}{2021}\natexlab{}.
\newblock \showarticletitle{Efficiently teaching an effective dense retriever
  with balanced topic aware sampling}. In \bibinfo{booktitle}{\emph{Proceedings
  of the 44th International ACM SIGIR Conference on Research and Development in
  Information Retrieval}}. \bibinfo{pages}{113--122}.
\newblock


\bibitem[Jiao et~al\mbox{.}(2020)]%
        {tinybert}
\bibfield{author}{\bibinfo{person}{Xiaoqi Jiao}, \bibinfo{person}{Yichun Yin},
  \bibinfo{person}{Lifeng Shang}, \bibinfo{person}{Xin Jiang},
  \bibinfo{person}{Xiao Chen}, \bibinfo{person}{Linlin Li},
  \bibinfo{person}{Fang Wang}, {and} \bibinfo{person}{Qun Liu}.}
  \bibinfo{year}{2020}\natexlab{}.
\newblock \showarticletitle{TinyBERT: Distilling BERT for Natural Language
  Understanding}.
\newblock \bibinfo{journal}{\emph{arXiv preprint arXiv:1909.10351}}
  (\bibinfo{year}{2020}).
\newblock


\bibitem[Lin et~al\mbox{.}(2022)]%
        {lin2022fostering}
\bibfield{author}{\bibinfo{person}{Jimmy Lin}, \bibinfo{person}{Daniel Campos},
  \bibinfo{person}{Nick Craswell}, \bibinfo{person}{Bhaskar Mitra}, {and}
  \bibinfo{person}{Emine Yilmaz}.} \bibinfo{year}{2022}\natexlab{}.
\newblock \showarticletitle{Fostering Coopetition While Plugging Leaks: The
  Design and Implementation of the MS MARCO Leaderboards}.
\newblock  (\bibinfo{year}{2022}).
\newblock


\bibitem[Lin et~al\mbox{.}(2021a)]%
        {lin2021pyserini}
\bibfield{author}{\bibinfo{person}{Jimmy Lin}, \bibinfo{person}{Xueguang Ma},
  \bibinfo{person}{Sheng-Chieh Lin}, \bibinfo{person}{Jheng-Hong Yang},
  \bibinfo{person}{Ronak Pradeep}, {and} \bibinfo{person}{Rodrigo Nogueira}.}
  \bibinfo{year}{2021}\natexlab{a}.
\newblock \showarticletitle{Pyserini: An Easy-to-Use Python Toolkit to Support
  Replicable IR Research with Sparse and Dense Representations}.
\newblock \bibinfo{journal}{\emph{arXiv preprint arXiv:2102.10073}}
  (\bibinfo{year}{2021}).
\newblock


\bibitem[Lin et~al\mbox{.}(2020)]%
        {lin2020pretrained}
\bibfield{author}{\bibinfo{person}{Jimmy Lin}, \bibinfo{person}{Rodrigo
  Nogueira}, {and} \bibinfo{person}{Andrew Yates}.}
  \bibinfo{year}{2020}\natexlab{}.
\newblock \showarticletitle{Pretrained transformers for text ranking: Bert and
  beyond}.
\newblock \bibinfo{journal}{\emph{arXiv preprint arXiv:2010.06467}}
  (\bibinfo{year}{2020}).
\newblock


\bibitem[Lin et~al\mbox{.}(2021b)]%
        {lin_2021}
\bibfield{author}{\bibinfo{person}{Xi~Victoria Lin}, \bibinfo{person}{Todor
  Mihaylov}, \bibinfo{person}{Mikel Artetxe}, \bibinfo{person}{Tianlu Wang},
  \bibinfo{person}{Shuohui Chen}, \bibinfo{person}{Daniel Simig},
  \bibinfo{person}{Myle Ott}, \bibinfo{person}{Naman Goyal},
  \bibinfo{person}{Shruti Bhosale}, \bibinfo{person}{Jingfei Du},
  \bibinfo{person}{Ramakanth Pasunuru}, \bibinfo{person}{Sam Shleifer},
  \bibinfo{person}{Punit~Singh Koura}, \bibinfo{person}{Vishrav Chaudhary},
  \bibinfo{person}{Brian O'Horo}, \bibinfo{person}{Jeff Wang},
  \bibinfo{person}{Luke Zettlemoyer}, \bibinfo{person}{Zornitsa Kozareva},
  \bibinfo{person}{Mona Diab}, \bibinfo{person}{Veselin Stoyanov}, {and}
  \bibinfo{person}{Xian Li}.} \bibinfo{year}{2021}\natexlab{b}.
\newblock \showarticletitle{Few-shot Learning with Multilingual Language
  Models}.
\newblock \bibinfo{journal}{\emph{arXiv preprint arXiv:2112.10668}}
  (\bibinfo{year}{2021}).
\newblock


\bibitem[Lu et~al\mbox{.}(2021)]%
        {lu-etal-2021-less}
\bibfield{author}{\bibinfo{person}{Shuqi Lu}, \bibinfo{person}{Di He},
  \bibinfo{person}{Chenyan Xiong}, \bibinfo{person}{Guolin Ke},
  \bibinfo{person}{Waleed Malik}, \bibinfo{person}{Zhicheng Dou},
  \bibinfo{person}{Paul Bennett}, \bibinfo{person}{Tie-Yan Liu}, {and}
  \bibinfo{person}{Arnold Overwijk}.} \bibinfo{year}{2021}\natexlab{}.
\newblock \showarticletitle{Less is More: Pretrain a Strong {S}iamese Encoder
  for Dense Text Retrieval Using a Weak Decoder}. In
  \bibinfo{booktitle}{\emph{Proceedings of the 2021 Conference on Empirical
  Methods in Natural Language Processing}}. \bibinfo{publisher}{Association for
  Computational Linguistics}, \bibinfo{address}{Online and Punta Cana,
  Dominican Republic}, \bibinfo{pages}{2780--2791}.
\newblock
\urldef\tempurl%
\url{https://doi.org/10.18653/v1/2021.emnlp-main.220}
\showDOI{\tempurl}


\bibitem[MacAvaney et~al\mbox{.}(2019)]%
        {macavaney2019cedr}
\bibfield{author}{\bibinfo{person}{Sean MacAvaney}, \bibinfo{person}{Andrew
  Yates}, \bibinfo{person}{Arman Cohan}, {and} \bibinfo{person}{Nazli
  Goharian}.} \bibinfo{year}{2019}\natexlab{}.
\newblock \showarticletitle{CEDR: Contextualized embeddings for document
  ranking}. In \bibinfo{booktitle}{\emph{Proceedings of the 42nd International
  ACM SIGIR Conference on Research and Development in Information Retrieval}}.
  \bibinfo{pages}{1101--1104}.
\newblock


\bibitem[Muennighoff(2022)]%
        {SGPT}
\bibfield{author}{\bibinfo{person}{Niklas Muennighoff}.}
  \bibinfo{year}{2022}\natexlab{}.
\newblock \showarticletitle{SGPT: GPT Sentence Embeddings for Semantic Search}.
\newblock \bibinfo{journal}{\emph{arXiv preprint arXiv:2202.08904}}
  (\bibinfo{year}{2022}).
\newblock


\bibitem[Mukherjee and Awadallah(2020)]%
        {xtremedistil}
\bibfield{author}{\bibinfo{person}{Subhabrata Mukherjee} {and}
  \bibinfo{person}{Ahmed Awadallah}.} \bibinfo{year}{2020}\natexlab{}.
\newblock \showarticletitle{XtremeDistil: Multi-stage Distillation for Massive
  Multilingual Models}.
\newblock \bibinfo{journal}{\emph{arXiv preprint arXiv:2004.05686}}
  (\bibinfo{year}{2020}).
\newblock


\bibitem[Ni et~al\mbox{.}(2021)]%
        {dual_2021}
\bibfield{author}{\bibinfo{person}{Jianmo Ni}, \bibinfo{person}{Chen Qu},
  \bibinfo{person}{Jing Lu}, \bibinfo{person}{Zhuyun Dai},
  \bibinfo{person}{Gustavo~Hernández Ábrego}, \bibinfo{person}{Ji Ma},
  \bibinfo{person}{Vincent~Y. Zhao}, \bibinfo{person}{Yi Luan},
  \bibinfo{person}{Keith~B. Hall}, \bibinfo{person}{Ming-Wei Chang}, {and}
  \bibinfo{person}{Yinfei Yang}.} \bibinfo{year}{2021}\natexlab{}.
\newblock \showarticletitle{Large Dual Encoders Are Generalizable Retrievers}.
\newblock \bibinfo{journal}{\emph{arXiv preprint arXiv:2112.07899}}
  (\bibinfo{year}{2021}).
\newblock


\bibitem[Nogueira and Cho(2019)]%
        {nogueira2019passage}
\bibfield{author}{\bibinfo{person}{Rodrigo Nogueira} {and}
  \bibinfo{person}{Kyunghyun Cho}.} \bibinfo{year}{2019}\natexlab{}.
\newblock \showarticletitle{Passage Re-ranking with BERT}.
\newblock \bibinfo{journal}{\emph{arXiv preprint arXiv:1901.04085}}
  (\bibinfo{year}{2019}).
\newblock


\bibitem[Nogueira et~al\mbox{.}(2020)]%
        {nogueira2020document}
\bibfield{author}{\bibinfo{person}{Rodrigo Nogueira}, \bibinfo{person}{Zhiying
  Jiang}, \bibinfo{person}{Ronak Pradeep}, {and} \bibinfo{person}{Jimmy Lin}.}
  \bibinfo{year}{2020}\natexlab{}.
\newblock \showarticletitle{Document Ranking with a Pretrained
  Sequence-to-Sequence Model}. In \bibinfo{booktitle}{\emph{Proceedings of the
  2020 Conference on Empirical Methods in Natural Language Processing:
  Findings}}. \bibinfo{pages}{708--718}.
\newblock


\bibitem[Pradeep et~al\mbox{.}(2020)]%
        {pradeep5h2oloo}
\bibfield{author}{\bibinfo{person}{Ronak Pradeep}, \bibinfo{person}{Xueguang
  Ma}, \bibinfo{person}{Xinyu Zhang}, \bibinfo{person}{Hang Cui},
  \bibinfo{person}{Ruizhou Xu}, \bibinfo{person}{Rodrigo Nogueira},
  \bibinfo{person}{Jimmy~J. Lin}, {and} \bibinfo{person}{David~R. Cheriton}.}
  \bibinfo{year}{2020}\natexlab{}.
\newblock \showarticletitle{H2oloo at TREC 2020: When all you got is a
  hammer... Deep Learning, Health Misinformation, and Precision Medicine}. In
  \bibinfo{booktitle}{\emph{TREC}}.
\newblock


\bibitem[Qu et~al\mbox{.}(2021)]%
        {qu2021rocketqa}
\bibfield{author}{\bibinfo{person}{Yingqi Qu}, \bibinfo{person}{Yuchen Ding},
  \bibinfo{person}{Jing Liu}, \bibinfo{person}{Kai Liu},
  \bibinfo{person}{Ruiyang Ren}, \bibinfo{person}{Wayne~Xin Zhao},
  \bibinfo{person}{Daxiang Dong}, \bibinfo{person}{Hua Wu}, {and}
  \bibinfo{person}{Haifeng Wang}.} \bibinfo{year}{2021}\natexlab{}.
\newblock \showarticletitle{RocketQA: An Optimized Training Approach to Dense
  Passage Retrieval for Open-Domain Question Answering}. In
  \bibinfo{booktitle}{\emph{Proceedings of the 2021 Conference of the North
  American Chapter of the Association for Computational Linguistics: Human
  Language Technologies}}. \bibinfo{pages}{5835--5847}.
\newblock


\bibitem[Raffel et~al\mbox{.}(2020)]%
        {raffel2020t5}
\bibfield{author}{\bibinfo{person}{Colin Raffel}, \bibinfo{person}{Noam
  Shazeer}, \bibinfo{person}{Adam Roberts}, \bibinfo{person}{Katherine Lee},
  \bibinfo{person}{Sharan Narang}, \bibinfo{person}{Michael Matena},
  \bibinfo{person}{Yanqi Zhou}, \bibinfo{person}{Wei Li}, {and}
  \bibinfo{person}{Peter~J. Liu}.} \bibinfo{year}{2020}\natexlab{}.
\newblock \showarticletitle{Exploring the Limits of Transfer Learning with a
  Unified Text-to-Text Transformer}.
\newblock \bibinfo{journal}{\emph{Journal of Machine Learning Research}}
  \bibinfo{volume}{21}, \bibinfo{number}{140} (\bibinfo{year}{2020}),
  \bibinfo{pages}{1--67}.
\newblock
\urldef\tempurl%
\url{http://jmlr.org/papers/v21/20-074.html}
\showURL{%
\tempurl}


\bibitem[Rajpurkar et~al\mbox{.}(2018)]%
        {squad2018}
\bibfield{author}{\bibinfo{person}{Pranav Rajpurkar}, \bibinfo{person}{Robin
  Jia}, {and} \bibinfo{person}{Percy Liang}.} \bibinfo{year}{2018}\natexlab{}.
\newblock \showarticletitle{Know What You Don't Know: Unanswerable Questions
  for SQuAD}.
\newblock \bibinfo{journal}{\emph{arXiv preprint arXiv:1806.03822}}
  (\bibinfo{year}{2018}).
\newblock


\bibitem[Ramponi and Plank(2020)]%
        {domain_adaptation}
\bibfield{author}{\bibinfo{person}{Alan Ramponi} {and} \bibinfo{person}{Barbara
  Plank}.} \bibinfo{year}{2020}\natexlab{}.
\newblock \showarticletitle{Neural Unsupervised Domain Adaptation in NLP---A
  Survey}.
\newblock \bibinfo{journal}{\emph{arXiv preprint arXiv:2006.00632}}
  (\bibinfo{year}{2020}).
\newblock


\bibitem[Reimers and Gurevych(2019)]%
        {reimers-2019-sentence-bert}
\bibfield{author}{\bibinfo{person}{Nils Reimers} {and} \bibinfo{person}{Iryna
  Gurevych}.} \bibinfo{year}{2019}\natexlab{}.
\newblock \showarticletitle{Sentence-BERT: Sentence Embeddings using Siamese
  BERT-Networks}. In \bibinfo{booktitle}{\emph{Proceedings of the 2019
  Conference on Empirical Methods in Natural Language Processing}}.
  \bibinfo{publisher}{Association for Computational Linguistics}.
\newblock
\urldef\tempurl%
\url{https://arxiv.org/abs/1908.10084}
\showURL{%
\tempurl}


\bibitem[Ren et~al\mbox{.}(2022)]%
        {dr_review}
\bibfield{author}{\bibinfo{person}{Ruiyang Ren}, \bibinfo{person}{Yingqi Qu},
  \bibinfo{person}{Jing Liu}, \bibinfo{person}{Wayne~Xin Zhao},
  \bibinfo{person}{Qifei Wu}, \bibinfo{person}{Yuchen Ding},
  \bibinfo{person}{Hua Wu}, \bibinfo{person}{Haifeng Wang}, {and}
  \bibinfo{person}{Ji-Rong Wen}.} \bibinfo{year}{2022}\natexlab{}.
\newblock \showarticletitle{A Thorough Examination on Zero-shot Dense
  Retrieval}.
\newblock \bibinfo{journal}{\emph{arXiv preprint arXiv:2204.12755}}
  (\bibinfo{year}{2022}).
\newblock


\bibitem[Roberts et~al\mbox{.}(2019)]%
        {Roberts2019OverviewOT}
\bibfield{author}{\bibinfo{person}{Kirk Roberts}, \bibinfo{person}{Dina
  Demner-Fushman}, \bibinfo{person}{E. Voorhees}, \bibinfo{person}{W. Hersh},
  \bibinfo{person}{Steven Bedrick}, \bibinfo{person}{Alexander~J. Lazar}, {and}
  \bibinfo{person}{S. Pant}.} \bibinfo{year}{2019}\natexlab{}.
\newblock \showarticletitle{Overview of the TREC 2019 Precision Medicine
  Track}.
\newblock \bibinfo{journal}{\emph{The ... text REtrieval conference : TREC.
  Text REtrieval Conference}}  \bibinfo{volume}{26} (\bibinfo{year}{2019}).
\newblock


\bibitem[Rosa et~al\mbox{.}(2021)]%
        {icail_2021}
\bibfield{author}{\bibinfo{person}{Guilherme~Moraes Rosa},
  \bibinfo{person}{Ruan~Chaves Rodrigues}, \bibinfo{person}{Roberto Lotufo},
  {and} \bibinfo{person}{Rodrigo Nogueira}.} \bibinfo{year}{2021}\natexlab{}.
\newblock \showarticletitle{To Tune or Not To Tune? Zero-shot Models for Legal
  Case Entailment}.
\newblock \bibinfo{journal}{\emph{ICAIL’21, Eighteenth International
  Conference on Artificial Intelligence and Law, June 21–25, 2021, São
  Paulo, Brazil}} (\bibinfo{year}{2021}).
\newblock


\bibitem[SANH et~al\mbox{.}(2020)]%
        {distilbert}
\bibfield{author}{\bibinfo{person}{Victor SANH}, \bibinfo{person}{Lysandre
  DEBUT}, \bibinfo{person}{Julien CHAUMOND}, {and} \bibinfo{person}{Thomas
  WOLF}.} \bibinfo{year}{2020}\natexlab{}.
\newblock \showarticletitle{DistilBERT, a distilled version of BERT: smaller,
  faster, cheaper and lighter}.
\newblock \bibinfo{journal}{\emph{arXiv preprint arXiv:1910.01108}}
  (\bibinfo{year}{2020}).
\newblock


\bibitem[Sanh et~al\mbox{.}(2021)]%
        {sahn_2021}
\bibfield{author}{\bibinfo{person}{Victor Sanh}, \bibinfo{person}{Albert
  Webson}, \bibinfo{person}{Colin Raffel}, \bibinfo{person}{Stephen~H. Bach},
  \bibinfo{person}{Lintang Sutawika}, \bibinfo{person}{Zaid Alyafeai},
  \bibinfo{person}{Antoine Chaffin}, \bibinfo{person}{Arnaud Stiegler},
  \bibinfo{person}{Teven~Le Scao}, \bibinfo{person}{Arun Raja},
  \bibinfo{person}{Manan Dey}, \bibinfo{person}{M~Saiful Bari},
  \bibinfo{person}{Canwen Xu}, \bibinfo{person}{Urmish Thakker},
  \bibinfo{person}{Shanya~Sharma Sharma}, \bibinfo{person}{Eliza Szczechla},
  \bibinfo{person}{Taewoon Kim}, \bibinfo{person}{Gunjan Chhablani},
  \bibinfo{person}{Nihal Nayak}, \bibinfo{person}{Debajyoti Datta},
  \bibinfo{person}{Jonathan Chang}, \bibinfo{person}{Mike Tian-Jian Jiang},
  \bibinfo{person}{Han Wang}, \bibinfo{person}{Matteo Manica},
  \bibinfo{person}{Sheng Shen}, \bibinfo{person}{Zheng~Xin Yong},
  \bibinfo{person}{Harshit Pandey}, \bibinfo{person}{Rachel Bawden},
  \bibinfo{person}{Thomas Wang}, \bibinfo{person}{Trishala Neeraj},
  \bibinfo{person}{Jos Rozen}, \bibinfo{person}{Abheesht Sharma},
  \bibinfo{person}{Andrea Santilli}, \bibinfo{person}{Thibault Fevry},
  \bibinfo{person}{Jason~Alan Fries}, \bibinfo{person}{Ryan Teehan},
  \bibinfo{person}{Tali Bers}, \bibinfo{person}{Stella Biderman},
  \bibinfo{person}{Leo Gao}, \bibinfo{person}{Thomas Wolf}, {and}
  \bibinfo{person}{Alexander~M. Rush}.} \bibinfo{year}{2021}\natexlab{}.
\newblock \showarticletitle{Multitask Prompted Training Enables Zero-Shot Task
  Generalization}.
\newblock \bibinfo{journal}{\emph{arXiv preprint arXiv:2110.08207}}
  (\bibinfo{year}{2021}).
\newblock


\bibitem[Santhanam et~al\mbox{.}(2021)]%
        {santhanam2021colbertv2}
\bibfield{author}{\bibinfo{person}{Keshav Santhanam}, \bibinfo{person}{Omar
  Khattab}, \bibinfo{person}{Jon Saad-Falcon}, \bibinfo{person}{Christopher
  Potts}, {and} \bibinfo{person}{Matei Zaharia}.}
  \bibinfo{year}{2021}\natexlab{}.
\newblock \showarticletitle{Colbertv2: Effective and efficient retrieval via
  lightweight late interaction}.
\newblock \bibinfo{journal}{\emph{arXiv preprint arXiv:2112.01488}}
  (\bibinfo{year}{2021}).
\newblock


\bibitem[Sun et~al\mbox{.}(2018)]%
        {mobileBERT}
\bibfield{author}{\bibinfo{person}{Zhiqing Sun}, \bibinfo{person}{Hongkun Yu},
  \bibinfo{person}{Xiaodan Song}, \bibinfo{person}{Renjie Liu},
  \bibinfo{person}{Yiming Yang}, {and} \bibinfo{person}{Denny Zhou}.}
  \bibinfo{year}{2018}\natexlab{}.
\newblock \showarticletitle{MobileBERT: a Compact Task-Agnostic BERT for
  Resource-Limited Devices}.
\newblock \bibinfo{journal}{\emph{arXiv preprint arXiv:2004.02984}}
  (\bibinfo{year}{2018}).
\newblock


\bibitem[Thakur et~al\mbox{.}(2022)]%
        {thakur2022domain}
\bibfield{author}{\bibinfo{person}{Nandan Thakur}, \bibinfo{person}{Nils
  Reimers}, {and} \bibinfo{person}{Jimmy Lin}.}
  \bibinfo{year}{2022}\natexlab{}.
\newblock \showarticletitle{Domain Adaptation for Memory-Efficient Dense
  Retrieval}.
\newblock \bibinfo{journal}{\emph{arXiv preprint arXiv:2205.11498}}
  (\bibinfo{year}{2022}).
\newblock


\bibitem[Thakur et~al\mbox{.}(2021)]%
        {beir}
\bibfield{author}{\bibinfo{person}{Nandan Thakur}, \bibinfo{person}{Nils
  Reimers}, \bibinfo{person}{Andreas Rücklé}, \bibinfo{person}{Abhishek
  Srivastava}, {and} \bibinfo{person}{Iryna Gurevych}.}
  \bibinfo{year}{2021}\natexlab{}.
\newblock \showarticletitle{BEIR: A Heterogeneous Benchmark for Zero-shot
  Evaluation of Information Retrieval Models}.
\newblock \bibinfo{journal}{\emph{arXiv preprint arXiv:2104.08663}}
  (\bibinfo{year}{2021}).
\newblock


\bibitem[Voorhees(2004)]%
        {trec2004}
\bibfield{author}{\bibinfo{person}{Ellen~M. Voorhees}.}
  \bibinfo{year}{2004}\natexlab{}.
\newblock \showarticletitle{Overview of the TREC 2004 Robust Track}.
\newblock \bibinfo{journal}{\emph{Proceedings of the Thirteenth Text REtrieval
  Conference, TREC 2004, Gaithersburg, Maryland, November 16-19, 2004}}
  (\bibinfo{year}{2004}).
\newblock


\bibitem[Wang et~al\mbox{.}(2018)]%
        {glue}
\bibfield{author}{\bibinfo{person}{Alex Wang}, \bibinfo{person}{Amanpreet
  Singh}, \bibinfo{person}{Julian Michael}, \bibinfo{person}{Felix Hill},
  \bibinfo{person}{Omer Levy}, {and} \bibinfo{person}{Samuel~R. Bowman}.}
  \bibinfo{year}{2018}\natexlab{}.
\newblock \showarticletitle{GLUE: A Multi-Task Benchmark and Analysis Platform
  for Natural Language Understanding}.
\newblock \bibinfo{journal}{\emph{arXiv preprint arXiv:1804.07461}}
  (\bibinfo{year}{2018}).
\newblock


\bibitem[Wang et~al\mbox{.}(2021)]%
        {wang2021gpl}
\bibfield{author}{\bibinfo{person}{Kexin Wang}, \bibinfo{person}{Nandan
  Thakur}, \bibinfo{person}{Nils Reimers}, {and} \bibinfo{person}{Iryna
  Gurevych}.} \bibinfo{year}{2021}\natexlab{}.
\newblock \showarticletitle{GPL: Generative Pseudo Labeling for Unsupervised
  Domain Adaptation of Dense Retrieval}.
\newblock \bibinfo{journal}{\emph{arXiv preprint arXiv:2112.07577}}
  (\bibinfo{year}{2021}).
\newblock


\bibitem[Wang et~al\mbox{.}(2020)]%
        {minilm}
\bibfield{author}{\bibinfo{person}{Wenhui Wang}, \bibinfo{person}{Furu Wei},
  \bibinfo{person}{Li Dong}, \bibinfo{person}{Hangbo Bao}, \bibinfo{person}{Nan
  Yang}, {and} \bibinfo{person}{Ming Zhou}.} \bibinfo{year}{2020}\natexlab{}.
\newblock \showarticletitle{MiniLM: Deep Self-Attention Distillation for
  Task-Agnostic Compression of Pre-Trained Transformers}.
\newblock \bibinfo{journal}{\emph{arXiv preprint arXiv:2002.10957}}
  (\bibinfo{year}{2020}).
\newblock


\bibitem[Wei et~al\mbox{.}(2022)]%
        {wei2022finetuned}
\bibfield{author}{\bibinfo{person}{Jason Wei}, \bibinfo{person}{Maarten Bosma},
  \bibinfo{person}{Vincent Zhao}, \bibinfo{person}{Kelvin Guu},
  \bibinfo{person}{Adams~Wei Yu}, \bibinfo{person}{Brian Lester},
  \bibinfo{person}{Nan Du}, \bibinfo{person}{Andrew~M. Dai}, {and}
  \bibinfo{person}{Quoc~V Le}.} \bibinfo{year}{2022}\natexlab{}.
\newblock \showarticletitle{Finetuned Language Models are Zero-Shot Learners}.
  In \bibinfo{booktitle}{\emph{International Conference on Learning
  Representations}}.
\newblock
\urldef\tempurl%
\url{https://openreview.net/forum?id=gEZrGCozdqR}
\showURL{%
\tempurl}


\bibitem[Xin et~al\mbox{.}(2021)]%
        {xin2021zero}
\bibfield{author}{\bibinfo{person}{Ji Xin}, \bibinfo{person}{Chenyan Xiong},
  \bibinfo{person}{Ashwin Srinivasan}, \bibinfo{person}{Ankita Sharma},
  \bibinfo{person}{Damien Jose}, {and} \bibinfo{person}{Paul~N Bennett}.}
  \bibinfo{year}{2021}\natexlab{}.
\newblock \showarticletitle{Zero-Shot Dense Retrieval with Momentum Adversarial
  Domain Invariant Representations}.
\newblock \bibinfo{journal}{\emph{arXiv preprint arXiv:2110.07581}}
  (\bibinfo{year}{2021}).
\newblock


\bibitem[Zhan et~al\mbox{.}(2022)]%
        {dense_2022}
\bibfield{author}{\bibinfo{person}{Jingtao Zhan}, \bibinfo{person}{Xiaohui
  Xie}, \bibinfo{person}{Jiaxin Mao}, \bibinfo{person}{Yiqun Liu},
  \bibinfo{person}{Min Zhang}, {and} \bibinfo{person}{Shaoping Ma}.}
  \bibinfo{year}{2022}\natexlab{}.
\newblock \showarticletitle{Evaluating Extrapolation Performance of Dense
  Retrieval}.
\newblock \bibinfo{journal}{\emph{arXiv preprint arXiv:2204.11447}}
  (\bibinfo{year}{2022}).
\newblock


\bibitem[Zhang et~al\mbox{.}(2020)]%
        {zhang2020rapidly}
\bibfield{author}{\bibinfo{person}{Edwin Zhang}, \bibinfo{person}{Nikhil
  Gupta}, \bibinfo{person}{Rodrigo Nogueira}, \bibinfo{person}{Kyunghyun Cho},
  {and} \bibinfo{person}{Jimmy Lin}.} \bibinfo{year}{2020}\natexlab{}.
\newblock \showarticletitle{Rapidly Deploying a Neural Search Engine for the
  COVID-19 Open Research Dataset}. In \bibinfo{booktitle}{\emph{Proceedings of
  the 1st Workshop on NLP for COVID-19 at ACL 2020}}.
\newblock


\bibitem[Zhong et~al\mbox{.}(2021)]%
        {metalmm}
\bibfield{author}{\bibinfo{person}{Ruiqi Zhong}, \bibinfo{person}{Kristy Lee},
  {and} \bibinfo{person}{Dan~Klein Zheng~Zhang}.}
  \bibinfo{year}{2021}\natexlab{}.
\newblock \showarticletitle{Adapting Language Models for Zero-shot Learning by
  Meta-tuning on Dataset and Prompt Collections}.
\newblock \bibinfo{journal}{\emph{arXiv preprint arXiv:2104.04670}}
  (\bibinfo{year}{2021}).
\newblock


\end{thebibliography}

%\pagebreak 

\onecolumn
%\pagebreak 
\appendix

\section{Appendix}

In this section, we provide graphs that show the zero-shot effectiveness when scaling the parameter count for each dataset on the BEIR benchmark.

\begin{figure*}[h]
  \centering
  \includegraphics[width=19cm, height=19cm]{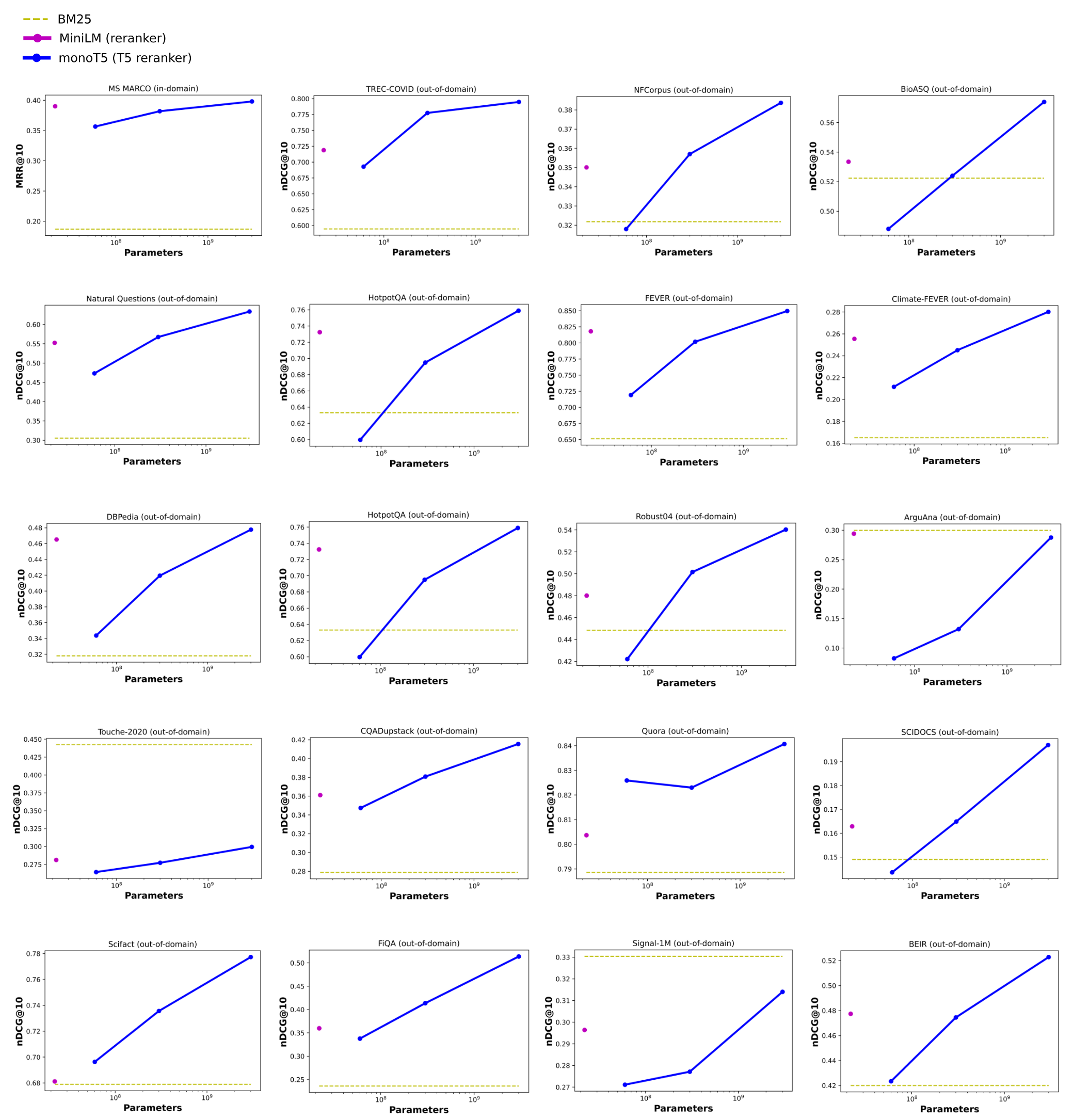}
  \caption{Model size vs effectiveness on in-domain (MS MARCO) vs out-of-domain (others) data. Effectiveness increase with respect to the number of model parameters only on out-of-domain datasets.}
  \label{fig:graphs}
\end{figure*}

\end{document}